\documentstyle[pra,aps]{revtex}
\begin{document}
\draft
\title{Rotational properties of trapped bosons}
\author{F.~Brosens and J.T. Devreese\cite{Author2},}
\address{Departement Natuurkunde, Universiteit Antwerpen (UIA),\\
Universiteitsplein 1, B-2610 Antwerpen}
\author{L. F. Lemmens,}
\address{Departement Natuurkunde, Universiteit Antwerpen (RUCA),\\
Groenenborgerlaan 171, B-2020 Antwerpen}
\date{September 2, revised version November 4, 1996.}
\maketitle

\begin{abstract}
Based on the Hellman-Feynman theorem it is shown that the average square
radius of a cloud of interacting bosons in a parabolic well can be derived
from their free energy. As an application, the temperature dependence of the
moment of inertia of non-interacting bosons in a parabolic trap is
determined as a function of the number of bosons. Well below the critical
condensation temperature, the Bose-Einstein statistics are found to
substantially reduce the moment of inertia of this system, as compared to a
gas of ``distinguishable'' particles in a parabolic well.
\end{abstract}


The study of a crowd of atoms in the same quantum state has gained great
importance since the recent observation of the Bose-Einstein condensation of
magnetically trapped gases of rubidium \cite{BEC1}, sodium \cite{BEC2} and
---albeit only with indirect evidence--- lithium \cite{BEC3}. According to
Stringari \cite{Stringari}, who draws a parallel with superfluid systems 
\cite{Baym}, the rotational properties of a confined Bose gas provide a
natural way to analyze the deviations from rigid motion due to condensation.
He suggested that condensation effects of magnetically trapped gases should
be observable in their moment of inertia, which he calculated approximately
from the linear response properties of a Bose gas of harmonic oscillators
without two-body interactions, relying on the semiclassical approximation 
\cite{DeGroot,Bagnato,Lewenstein} for the excited states.

Using the Hellman-Feynman theorem, we show in the present letter that the
moment of inertia of a confined Bose gas can be derived directly from the
dependence of its free energy on the strength of the confining parabolic
potential, even if two-body interactions are included. Indeed, for the
following Hamiltonian 
\begin{equation}
{\cal H} 
\begin{array}[t]{cccccc}
= & \bigskip \sum_{j=1}^N\frac{p_j^2}{2m} & + & \sum_{j<l}^Nv\left( \vec{r}%
_j-\vec{r}_l\right) & + & \frac m2\sum_j^N\left( \omega _z^2z_j^2+\omega
^2\left( x_j^2+y_j^2\right) \right) \\ 
= & {\cal H}_{kin} & + & {\cal H}_{int} & + & {\cal H}_{conf}
\end{array}
\end{equation}
it is straightforward to show that the confinement contribution ${\cal U}%
_{conf}\equiv \left\langle {\cal H}_{conf}\right\rangle $ to the internal
energy can be written as 
\begin{equation}
{\cal U}_{conf}=\frac 1{{\cal Z}}\text{Tr}\left( e^{-\beta {\cal H}}{\cal H}%
_{conf}\right) =\frac m2N\left( \omega _z^2\left\langle z^2\right\rangle
+\omega ^2\left\langle x^2+y^2\right\rangle \right) ,
\end{equation}
where ${\cal Z}$ is the partition function. Denoting the free energy by $%
{\cal F},$ the Hellman-Feynman theorem states that 
\begin{equation}
{\cal U}_{conf}=\omega _z^2\frac{\partial {\cal F}}{\partial \left( \omega
_z^2\right) }+\omega ^2\frac{\partial {\cal F}}{\partial \left( \omega
^2\right) },
\end{equation}
relating thereby the average square radius $\left\langle \rho
^2\right\rangle \equiv \left\langle x^2+y^2\right\rangle $ of the system in
the $xy$-plane (and hence its moment of inertia with respect to the $z$%
-axis) to partial derivatives of the free energy. This relation is valid
both for distinguishable and for identical particles. If the free energy
exhibits critical behavior, it will also be reflected in the average square
radius of the Bose gas and consequently in its rotational properties.

In order to estimate the moment of inertia quantitatively, we consider for
simplicity an ideal cloud of non-interacting bosons trapped in a parabolic
well characterized by a frequency $w.$ With $\mu $ denoting the chemical
potential and $\beta =1/kT$ where $k$ is Boltzmann's constant, and $T$ the
temperature, the partition function in the {\sl grand-canonical ensemble}
becomes 
\begin{equation}
\Xi _\mu =\prod_{\nu =0}^\infty \left( \frac 1{1-e^{-\beta \left( \epsilon
_\nu -\mu \right) }}\right) ^{\frac 12\left( \nu +1\right) \left( \nu
+2\right) },  \label{ksi_mu}
\end{equation}
\begin{equation}
\epsilon _\nu =\left( \nu +\frac 32\right) \hbar w;\quad \nu =0,1,2,\cdots ,
\end{equation}
where $\frac 12\left( \nu +1\right) \left( \nu +2\right) $ is the number of
different possibilities to occupy the $\nu ^{\text{th}}$ level. This
multiplicity depends on the symmetry and is considered here for the
isotropic case $\omega _z^2=\omega ^2$. For cylindrical symmetry, e.g., it
would be 1 along the symmetry axis, and $\left( \nu +1\right) $ in the plane
orthogonal to the symmetry axis.

The expression (\ref{ksi_mu}) is the isotropic limit of the grand canonical
partition function for a Bose gas of anisotropic oscillators \cite{Grossman}%
, from which the subsequent derivation can be generalized to the anisotropic
case $\omega _z^2\neq \omega ^2$. The subsequent discussion is limited to
the isotropic Bose gas. The ground state occupancy and the thermodynamic
properties including the anisotropic case have recently been analyzed in
several papers \cite{Grossman,Grossman2,Ketterle} and are currently also the
subject of some preprints \cite{Kirsten,Haugerud,Politzer}.

The average number of bosons $N=\frac 1\beta \frac \partial {\partial \mu }%
\ln \Xi _\mu $ and their internal energy $U=-\frac{\partial \ln \Xi _\mu }{%
\partial \beta }+N\mu $ in an isotropic parabolic confining potential are: 
\begin{equation}
N=\sum_{\nu =0}^\infty n_\nu ;\quad U=\sum_{\nu =0}^\infty n_\nu \epsilon
_\nu ;\quad n_\nu =\frac 12\left( \nu +1\right) \left( \nu +2\right) \frac{%
e^{-\beta \left( \epsilon _\nu -\mu \right) }}{1-e^{-\beta \left( \epsilon
_\nu -\mu \right) }},  \label{UandN}
\end{equation}
where $n_\nu $ is the occupation number of the $\nu ^{\text{th}}$ level.
Furthermore the Hellman-Feynman theorem leads to the average square radius
of the boson cloud 
\begin{equation}
\frac 12mN\left\langle \left. \vec{r}\right. ^2\right\rangle =-\frac 1\beta 
\frac d{d\left( w^2\right) }\ln \Xi _\mu =\frac 1{4w^2}\sum_{\nu =0}^\infty
\left( \nu +1\right) \left( \nu +2\right) \epsilon _\nu \frac{e^{-\beta
\left( \epsilon _\nu -\mu \right) }}{1-e^{-\beta \left( \epsilon _\nu -\mu
\right) }}=\frac 12\frac U{w^2}.
\end{equation}
Consequently, the averaged potential energy becomes $\frac 12%
mw^2N\left\langle \left. \vec{r}\right. ^2\right\rangle =\frac 12U,$ as
could have been anticipated from the virial theorem. The moment of inertia
relative to the $z$-axis is therefore given by: 
\begin{equation}
\Theta =mN\left\langle x^2+y^2\right\rangle =\frac 23\frac U{w^2},
\label{ThetaU}
\end{equation}
which is the basic result of the present letter. Critical behavior of the
internal energy (and consequently in the specific heat) of the model is
directly reflected in the rotational properties of the boson gas.

The evaluation of the internal energy for a given number $N$ of particles,
requires the knowledge of the chemical potential. By expressing the chemical
potential as a function of the number of particles in the ground state 
\begin{equation}
n_0=\frac{e^{-\beta \left( \frac 32\hbar w-\mu \right) }}{1-e^{-\beta \left( 
\frac 32\hbar w-\mu \right) }}\Longrightarrow e^{-\beta \left( \frac 32\hbar
w-\mu \right) }=\frac{n_0}{n_0+1},
\end{equation}
the total number of particles $N$ is obtained in terms of $n_0$:

\begin{equation}
N=\sum_{\nu =0}^\infty \frac 12\left( \nu +1\right) \left( \nu +2\right)
\left( \frac{n_0e^{-\beta \hbar w\nu }}{n_0+1-n_0e^{-\beta \hbar w\nu }}%
\right) .  \label{Nbad}
\end{equation}
However, as it turns out, this series is not very appropriate for numerical
treatment. A much more efficient and numerically well-convergent expression
results if each term in the denominator of (\ref{Nbad}) is expanded in
powers of $e^{-\beta \hbar w\nu },$ which allows to perform the summation
over $\nu .$ The same procedure can be used for the internal energy. The
results can be written in the following form, which is numerically stable: 
\begin{equation}
\begin{array}{lllll}
N & = & \sum_{\ell =1}^\infty \left( \frac{n_0}{n_0+1}\right) ^\ell \frac 1{%
\left( 1-e^{-\beta \hbar w\ell }\right) ^3} & = & n_0+\sum_{\ell =1}^\infty
\left( \frac{n_0}{n_0+1}\right) ^\ell \left( \frac 1{\left( 1-e^{-\beta
\hbar w\ell }\right) ^3}-1\right) ; \\ 
U & = & \frac 32\hbar w\sum_{\ell =1}^\infty \left( \frac{n_0}{n_0+1}\right)
^\ell \frac{1+e^{-\beta \hbar w\ell }}{\left( 1-e^{-\beta \hbar w\ell
}\right) ^4} & = & \frac 32\hbar w\left( N+2\sum_{\ell =1}^\infty \left( 
\frac{n_0}{n_0+1}\right) ^\ell \frac{e^{-\beta \hbar w\ell }}{\left(
1-e^{-\beta \hbar w\ell }\right) ^4}\right) .
\end{array}
\label{UNgood}
\end{equation}
In these expressions, $\ell $ is the length of a cycle in the cyclic
decomposition of the partition function\cite{Feynman} obtained from the
symmetrized density matrix for $N$ oscillators without two-body interactions.

The high temperature limit $\beta \rightarrow 0$ of the internal energy and
of the depletion of the condensate can now be analyzed directly. There is no
need anymore to make the continuum approximation by introducing a
parametrized form for the density of states \cite{Grossman}. In this limit,
and {\sl without making the continuum approximation,} $N-n_0$ becomes: 
\begin{equation}
\frac{kT}{\hbar w}\gg 1\Longrightarrow N-n_0\approx \sum_{\ell =1}^\infty
\left( \frac{n_0}{n_0+1}\right) ^\ell \left( \frac 1{\hbar ^3w^3\beta ^3}%
\frac 1{\ell ^3}+\frac 1{\hbar ^2w^2\beta ^2}\frac 3{2\ell ^2}+O\left( \frac %
1\beta \right) \right) ,
\end{equation}
and similarly for the internal energy $U$. For $n_0\gg 1,$ the summations
yield 
\begin{equation}
\begin{array}{l}
\frac{kT}{\hbar w}\gg 1 \\ 
n_0\gg 1
\end{array}
\Longrightarrow 
\begin{array}{l}
N-n_0=N\left( \frac T{T_c}\right) ^3\left( 1+\frac{3\zeta \left( 2\right) }{%
2\zeta \left( 3\right) }\frac{\hbar w}{kT}\right) +O\left( T\right) \\ 
\frac U{\frac 32\hbar wN}=1+2\left( \frac T{T_c}\right) ^3\left( \frac{kT}{%
\hbar w}\frac{\zeta \left( 4\right) }{\zeta \left( 3\right) }+1+O\left( 
\frac 1T\right) \right)
\end{array}
\text{ with }T_c=\frac{\hbar w}k\sqrt[3]{\frac N{\zeta \left( 3\right) }}.
\end{equation}

This result for $N-n_0$ has been derived (also {\sl without} the continuum
approximation) in \cite{Ketterle} and coincides with the result within an
improved continuum approximation as derived in \cite{Grossman} and \cite
{Ketterle}. The temperature expansion for the internal energy $U$ is
obtained in \cite{Grossman} using the continuum approximation.

For arbitrary temperature, the ground state occupancy $n_0$ and the internal
energy $U$ have to be evaluated numerically from (\ref{UNgood}). This can
efficiently be done by first solving for the quantity $\frac{n_0}{n_0+1}$
which is bracketed between 0 and 1. The resulting ground state occupancy
clearly exhibits the onset of a phase transition near the transition
temperature \cite{Lewenstein,Grossman,Grossman2,Haugerud,Politzer}.

Denoting by $\Theta _D$ the moment of inertia of the ``distinguishable''
oscillators (i.e. without Bose-Einstein statistics), we also calculated the
relative excess moment of inertia $\left( \Theta -\Theta _D\right) /\Theta
_D $ numerically. The results are shown in Fig. 1 as a function of the
reduced temperature for several values of $N.$ For $N$ of order 100 or more
the effect of the Bose statistics on the moment of inertia is very
pronounced, even reaching 75\% for $N=1000$ and 90\% for $N=10000$.

The Hellman-Feynman theorem is also applicable to energy estimates based on
the variational principle of quantum mechanics \cite{123,boundpol}, and it
can be used for other models than the isotropic gas of bosons in a parabolic
well as considered here. For instance, by applying the Hellman-Feynman
theorem using the internal energy from the semiclassical approximation \cite
{Bagnato}, one finds the same moment of inertia as was derived by Stringari 
\cite{Stringari} from the rotational response properties of this
semiclassical model. Note that the functions $Q\left( \eta \right)
=\int_0^\infty s^\eta /\left( \exp s-1\right) ds,$ used in equation (9--10)
and (13) of Ref. 
%TCIMACRO{\TeXButton{onlinecite:Stringari}{\onlinecite{Stringari}}}
%BeginExpansion
\onlinecite{Stringari}%
%EndExpansion
, are related to the Riemann $\zeta $ functions: $Q\left( 2\right) =2\zeta
\left( 3\right) $ and $Q\left( 3\right) =6\zeta \left( 4\right) ,$ and are a
special case of the well-known Bose-Einstein functions $g_n\left( z\right)
=\sum_{j=1}^\infty z^j/j^n,$ the properties of which are described in
standard textbooks, e.g., in Ref. 
%TCIMACRO{\TeXButton{onlinecite:Pathria}{\onlinecite{Pathria}}}
%BeginExpansion
\onlinecite{Pathria}%
%EndExpansion
.

It should be emphasized that the Hellman-Feynman theorem is valid in the
canonical as well as in the grand canonical ensemble. Because the internal
energy in the grand canonical ensemble with an average number of $N$ bosons
equals the internal energy in the canonical ensemble with a fixed number of $%
N$ bosons \cite{BDLSSC}, also the moments of inertia of both ensembles are
the same.

Having elucidated that the moment of inertia can be derived from the
confinement energy in some models for the ideal or almost ideal gas, we did
not discuss how quantitative results can be obtained in the presence of
two-body interactions. Well aware of this limitation in the present letter,
we like to point out that the inclusion of harmonic two-body interactions
presents no {\sl conceptual} difficulty. Their influence on the partition
function and on the thermodynamical properties can be and has been obtained 
\cite{BDLSSC} with a technique which we developed for treating the path
integral for identical particles using the Feynman-Kac functional and
invoking an appropriate linear combination of boson and fermion diffusion
processes \cite{LBD,BDL,LBDPR}.

In summary, the partition function of an ideal Bose gas in a parabolic trap
is has been obtained in the present letter, and the ground state occupancy $%
n_0,$ the internal energy $U$ and the moment of inertia $\Theta $ have been
derived. The rotational properties of the trapped gas are related to the
internal energy through the Hellman-Feynman theorem. In the isotropic model
we found the relation $\Theta =\frac 23\frac U{w^2},$ implying that the
critical behavior of the internal energy will manifest itself
macroscopically in the moment of inertia. The semi-classical approximation%
\cite{DeGroot,Bagnato,Stringari} turns out to provide the correct asymptotic
limits of this exactly soluble model. Moreover, the path integral method
allows to study this model as the zeroth order approximation in examining
the role of the two-body interactions, using many-body diffusion and the
Feynman-Kac functional.

\acknowledgments
Part of this work is performed in the framework of the NFWO projects No.
2.0093.91, 2.0110.91, G. 0287.95 and WO.073.94N (Wetenschappelijke
Onderzoeksgemeenschap, Scientific Research Community of the NFWO on
``Low-Dimensional Systems''), and in the framework of the European Community
Program Human Capital and Mobility through contracts no. CHRX-CT93-0337 and
CHRX-CT93-0124. One of the authors (F.B.) acknowledges the National Fund for
Scientific Research for financial support.

\bigskip {\bf Figure Captions}

\begin{description}
\item[Fig. 1]  Relative excess moment of inertia $\left( \Theta -\Theta
_D\right) /\Theta _D$ with respect to the $z$-axis as a function of $T/T_c$
for $N=$ 10, 100, 1000 and 10000.
\end{description}

%TCIMACRO{\TeXButton{end{document}}{\end{document}}}
%BeginExpansion
\end{document}